# Faster Replacement Paths[*]

Virginia Vassilevska Williams[†]

Computer Science Division, University of California, Berkeley.

**Abstract**

The replacement paths problem for directed graphs is to find for given nodes $s$ and $t$ and every edge $e$ on the shortest path between them, the shortest path between $s$ and $t$ which avoids $e$. For unweighted directed graphs on $n$ vertices, the best known algorithm runtime was $\tilde{O}(n^{2.5})$ by Roditty and Zwick. For graphs with integer weights in $\{-M, \ldots, M\}$, Weimann and Yuster recently showed that one can use fast matrix multiplication and solve the problem in $O(Mn^{2.584})$ time, a runtime which would be $O(Mn^{2.33})$ if the exponent $\omega$ of matrix multiplication is 2.

We improve both of these algorithms. Our new algorithm also relies on fast matrix multiplication and runs in $O(Mn^\omega \operatorname{poly} \log n)$ time if $\omega > 2$ and $O(Mn^{2+\varepsilon})$ for any $\varepsilon > 0$ if $\omega = 2$. Our result shows that, at least for small integer weights, the replacement paths problem in directed graphs may be easier than the related all pairs shortest paths problem in directed graphs, as the current best runtime for the latter is $\Omega(n^{2.5})$ time even if $\omega = 2$.

## 1 Introduction

In the replacement paths problem, one is given a graph and a shortest path $P$ between two vertices $s$ and $t$, and one needs to return for every edge on $P$, the shortest path between $s$ and $t$ which avoids $e$. The typical motivation for the problem is that links in a network can fail and a backup path between important vertices would be useful. Replacement paths are useful also in contexts in which one may wish to satisfy other constraints beyond short length [15]. For instance, in biological sequence alignment [4] replacement paths are useful in determining which pieces of an alignment are most important. The replacement paths problem is also used in the computation of Vickrey Prices of edges that are owned by selfish agents in a network [18, 12], and in finding the $k$ shortest simple paths between two nodes [15, 23, 19, 20].

The replacement paths problem for undirected graphs can be solved very efficiently: Malik et al. [16] gave an $\tilde{O}(m)$[1] algorithm for graphs on $m$ edges; Nardelli et al. [17] used Thorup's linear time algorithm for single source shortest paths to improve the runtime to $O(m\alpha(n))$ in the word-RAM model of computation. The best algorithm for the problem in directed graphs with arbitrary edge weights is by Gotthilf and Lewenstein [11] and runs in $O(mn + n^2 \log \log n)$ time, where $m$ and $n$ are the number of edges and vertices respectively. For dense graphs, nothing much better than cubic time is known.[2] Recently, Vassilevska-Williams and Williams [22] showed that the general replacement paths problem in directed graphs is equivalent to the all pairs shortest paths problem (APSP), under subcubic reductions. This essentially means that either both problems admit truly subcubic algorithms, or neither of them does. It is worth pointing out that this apparent cubic time barrier is only due to the wish to compute the replacement distances *exactly*. In contrast, Bernstein [2] recently showed that there is an almost *linear* time approximation scheme for replacement paths.

For *unweighted* directed graphs, Roditty and Zwick [20] gave a randomized combinatorial algorithm which computes replacement paths in $\tilde{O}(m\sqrt{n})$ time. In their recent FOCS paper, Weimann and Yuster [21] were the first to apply fast matrix multiplication techniques to the problem. Using the current best algorithm for matrix multiplication [6], Weimann and Yuster obtain a randomized algorithm for replacement paths which runs in $O(Mn^{2.584})$ for $n$ node graphs with integer weights in $\{-M, \ldots, M\}$. It has the advantage that it can also be used to construct a distance sensitivity oracle. Furthermore, if the exponent $\omega$ of matrix multiplication is actually 2, their runtime would be $O(Mn^{2.33})$ which would improve on Roditty and Zwick's $O(n^{2.5})$ runtime for unweighted graphs.

---

[*]Supported by a Computing Innovations Fellowship.

[†]Email: `virgi@eecs.berkeley.edu`

[1]The notation $\tilde{O}(f(n))$ means $O(f(n) \operatorname{poly} \log(n))$.

[2]Polylogarithmic improvements are possible. For example, Chan's [5] $O(n^3/\log^2 n)$ algorithm for all pairs shortest paths can be converted to one for replacement paths.

In this work we reconsider the replacement paths problem for dense directed graphs with weights in $\{-M, \ldots, M\}$. We give a simple reduction from the problem to all pairs shortest paths (APSP) and by applying Zwick's algorithm [26], we obtain an improved *deterministic* algorithm for replacement paths.

**Theorem 1.** *There is a deterministic algorithm for the replacement paths problem in directed graphs with weights in $\{-M, \ldots, M\}$ which runs in $\tilde{O}(M^{0.681} n^{2.575})$ time.*

This runtime bound improves on Weimann and Yuster's randomized algorithm (although if $\omega = 2$, Weimann and Yuster's bound would be lower). We then combine the ideas from the APSP reduction with a divide and conquer approach to design a faster randomized algorithm for replacement paths.

**Theorem 2.** *The replacement paths problem in a directed graph with integer weights in $\{-M, \ldots, M\}$ can be computed by a randomized algorithm in $O(M^{0.832} n^{2.481})$ time.*

Finally, we use a recursive approach on top of the ideas used in proving Theorem 2 to improve the runtime even further:

**Theorem 3.** *Let $\omega < 2.376$ [6] be the exponent for matrix multiplication. The replacement paths problem in a directed graph with integer weights in $\{-M, \ldots, M\}$ can be computed by a randomized algorithm in $O(Mn^{\omega} \text{poly} \log n)$ time, if $\omega > 2$, or in $O(Mn^{2+\varepsilon})$ for every $\varepsilon > 0$ if $\omega = 2$.*

Because of prior reductions between the replacement paths problem and the $k$ shortest simple paths problem [20] we obtain the following corollary.

**Corollary 1.** *The $k$ shortest simple paths in a directed graph with edge weigths in $\{-M, \ldots, M\}$ can be computed in randomized $O(k \cdot Mn^{\omega} \text{poly} \log n)$ time.*

Theorem 3 improves on Roditty and Zwick's $O(n^{2.5})$ runtime. It also improves the range of $M$ for which there is a subcubic algorithm for the problem: the previous range of Weimann and Yuster's algorithm was for $M = O(n^{0.416})$, and our algorithms are subcubic for any $M = O(n^{0.623})$. Our result also shows that, at least for small integer weights, the replacement paths problem in directed graphs might actually be easier than APSP in directed graphs. The current best algorithm for APSP in unweighted directed graphs is by Zwick [26]; it uses fast matrix multiplication and would run in $\Omega(n^{2.5})$ time even if $\omega = 2$. Furthermore, improving on $O(n^{\omega})$ for replacement paths in directed unweighted graphs would likely require radically new techniques, as the problem is very related to Boolean matrix multiplication.

**Other related work.** The replacement paths problem is closely related to the problem of finding the second shortest path between two nodes, and in general to the $k$ *shortest paths* problem. For directed graphs with $m$ edges and $n$ vertices and nonnegative edge weights, Eppstein [8] gave an algorithm which returns the $k$ shortest paths between two given nodes $s$ and $t$ in time $O(m + n \log n + k)$. The paths that Eppstein's algorithm returns, however, may not be simple. When the $k$ shortest paths are required to be simple, the fastest known algorithms for the problem use algorithms for the replacement paths problem. Roditty and Zwick [20] showed that the $k$ simple shortest paths problem can be reduced to $O(k)$ computations of the second shortest simple path. For unweighted graphs, both the replacement paths problem and the second shortest simple path problem are closely related to Boolean matrix multiplication; for general directed graphs with arbitrary weights, both problems are equivalent to all pairs shortest paths under subcubic reductions [22]. Hershberger et al. [13] showed that for directed graphs with arbitrary edge weights, the replacement paths problem requires $\Omega(n^{2.5})$ time in the path-comparison model of computation of Karger et al. [14].

**Preliminaries.** All graphs in this paper have $n$ vertices, unless otherwise stateds. They are directed and weakly connected. Each graph $G = (V, E)$ is equipped with a weight function, $\ell : E \to \{-M, \ldots, M\}$ where $M$ is a positive integer. The *length* or *weight* of a path $v_1 \to v_2 \to \ldots \to v_t$ in $G$ is just $\sum_{i=1}^{t-1} \ell(v_i, v_{i+1})$. A *shortest* path between two nodes is just a path between them which minimizes the length function.

Given a path $P = v_1 \to v_2 \to \ldots \to v_t$ in $G$, let $d_P(v_i, v_j) = \sum_{k=i}^{j-1} \ell(v_k, v_{k+1})$, *i.e.* the length of the subpath of $P$ from $v_i$ to $v_j$.

For an integer $n > 0$, let $[n] = \{1, 2, \ldots, n\}$. Let $A$ be an $m \times n$ matrix and $B$ be an $n \times p$ matrix, where $A$ and $B$ have entries from $\mathbb{Z} \cup \{\infty\}$. Then their *distance* product $A \star B$ is defined as

$$(A \star B)[i, j] = \min_{k \in [n]} (A[i, k] + B[k, j]), \ \forall i \in [m], j \in [p].$$

Finally, let $\omega$ be the infimum of all numbers such that there exists a constant $n_0$ so that for all $n > n_0$, the product of two $n \times n$ integer matrices can be computed in $\tilde{O}(n^\omega)$ time.

**Replacement paths problem.** Given $G = (V, E)$ with edge weights in $\{-M, \ldots, M\}$ and $s, t \in V$, the task is to compute the shortest path $P$ from $s$ to $t$ in $G$ and for every edge $e \in P$, the shortest path in $G \setminus \{e\}$. It is known how to compute the actual shortest path $P$ efficiently. For instance, when the weights are all nonnegative, one can just use Dijkstra's algorithm [7] with Fibonacci heaps [9] and find $P$ in $O(m + n \log n)$ time which is linear for dense graphs. When the weights can be negative, Goldberg's algorithm [10] shows how to find $P$ or detect a negative cycle in the graph in $O(m\sqrt{n} \operatorname{poly} \log M)$ time. Yuster and Zwick's [24] algorithm relies on fast matrix multiplication and finds $P$ in $O(Mn^\omega)$ time; using Coppersmith and Winograd's [6] bound on $\omega$, their running time is $O(Mn^{2.376})$ which is better than Goldberg's runtime for dense graphs and small $M$. Relying on the existing subcubic algorithms, we can from now on assume that $P$ is given as part of the input.

The naive way to compute the replacement paths is to remove each edge $e \in P$ in turn and compute the shortest path in $G \setminus \{e\}$ from scratch. This approach is however unnecessarily time intensive – the shortest paths computations share a lot of information.

## 2 Two Simple Algorithms

The notion of a *detour* is very important for the replacement paths problem. Let $P = \{s = v_1 \to v_2 \to \ldots \to v_k = t\}$ be the shortest path between $s$ and $t$ in $G$. Let $E(P)$ denote the set $\{(v_i, v_{i+1})\}_{i \in [k-1]}$ of all edges on $P$. For $j > i$, a detour $D(v_i, v_j)$ between $v_i$ and $v_j$ is a simple path from $v_i$ to $v_j$ which does not contain any other nodes of $P$. A detour $D(v_j, v_k)$ is said to *circumvent* edge $(v_i, v_{i+1})$ if $j \leq i$ and $i + 1 \leq k$. It is well known (see e.g. [2], Lemma 2.1) that for any edge $e = (v_i, v_{i+1}) \in E(P)$, the shortest path between $s$ and $t$ in $G \setminus \{e\}$ is exactly the minimum out of all paths of the form

$$s \to v_2 \to \ldots \to v_j \odot D(v_j, v_k) \odot v_k \to \ldots \to t,$$

where $j \leq i$ and $i + 1 \leq k$.

From the above discussion, we see that in order to find all replacement paths, it is sufficient to compute all the detours between nodes in $P$. After all detours are found, we can compute in $O(n^2)$ extra time for every detour $D(v_i, v_j)$, the length $\ell(D(v_i, v_j))$ of the path $s \to \ldots \to v_j \odot D(v_j, v_k) \odot v_k \to \ldots \to t$ which it defines. This can be done by first computing the weight of each subpath $s \to \ldots \to v_i$ for all $i$ and the weight of each subpath $v_j \to \ldots \to t$, via dynamic programming; then $\ell(D(v_i, v_j))$ can be computed in constant time by adding the weights of the two paths and that of the detour.

After this we can sort the detours in nondecreasing order according to $\ell(\cdot)$ in $O(n^2 \log n)$ time. We can then store all nodes $v_i$ of $P$ in a successor search data structure $T$ (e.g. any balanced binary search tree), sorted according to $i$, their position in $P$. To process the current shortest detour $D(v_i, v_j)$ in the sorted order, find the successor $v_x$ of $v_i$ in $T$, and if $x \leq j - 1$, record that the shortest replacement path length[3] for $(v_x, v_{x+1})$ is $\ell(D(v_i, v_j))$. Then remove $v_x$ from $T$ and process the rest of the elements in $T$ which are between $v_i$ and $v_j$. Collecting all the results from the detour information only takes an additional $O(n^2 \log n)$ time. The algorithms below show how to compute the detours.

---

[3] The actual path can also be stored, as usual, with a matrix of successors.

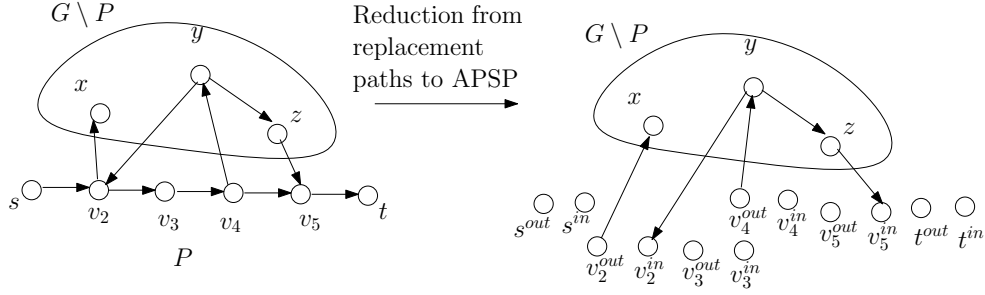

Figure 1: An example of the reduction from replacement paths to APSP in the case of an unweighted graph.

**Algorithm 1: A reduction to APSP.** We show that the detours can be computed using a reduction to APSP. The possibility of such a reduction was mentioned in the work of Bhosle and Gonzalez [3]; here we make it explicit and show that it can be used to further improve the known runtime bounds.

For every node $v_i$ of $P$, create two copies $v_i^{in}$ and $v_i^{out}$. Create a new graph $G'$ by taking $G \setminus P \cup \{v_i^{in}, v_i^{out}\}_{i \in [k]}$. For every edge $(v_i, u)$ for $u \in G \setminus P$, add an edge $(v_i^{out}, u)$ of the same weight in $G'$. Similarly, for every edge $(u, v_i)$ for $u \in G \setminus P$, add an edge $(u, v_i^{in})$ of the same weight in $G'$. $G'$ is essentially $G$ with the edges of $P$ removed, except that each node of $P$ is split into two. See Figure 1 for an example conversion from $G$ to $G'$. Now compute all pairs shortest paths in $G'$. The shortest path between $v_i^{out}$ and $v_j^{in}$ is exactly the optimal detour $D(v_i, v_j)$ in $G$. Thus with one call to an APSP algorithm, and $O(n^2 \log n)$ extra time we obtain an algorithm for replacement paths.

**Theorem 4** (Zwick's APSP [26]). *The APSP problem in a directed graph with integer edge weights in $\{-M, \ldots, M\}$ can be computed in $\tilde{O}(M^{1/(4-\omega)} n^{2+1/(4-\omega)})$ time, where $\omega < 2.376$ is the exponent of matrix multiplication. If rectangular matrix multiplication is used, then the running time is $O(M^{0.681} n^{2.575})$.*

Applying Zwick's [26] algorithm for APSP from Theorem 4 with weights in $\{-M, \ldots, M\}$ we prove Theorem 1.

**Reminder of Theorem 1.** *There is a deterministic algorithm for the replacement paths problem in directed graphs with weights in $\{-M, \ldots, M\}$ which runs in $\tilde{O}(M^{0.681} n^{2.575})$ time.*

**Algorithm 2: Divide and Conquer.** This second algorithm has a worse running time, but illustrates the fact that the problem lends itself to some partitioning. Consider a subpath $P'$ of $P$ lying between nodes $v_x$ and $v_y$. Let $V(P')$ be the nodes of $P'$ except for $v_x$ and $v_y$. Consider an edge $(v_i, v_{i+1})$ in $P'$. Then, the best detour $D(v_j, v_k)$ which circumvents $(v_i, v_{i+1})$ either has both $v_j$ and $v_k$ in $V(P')$, or $v_j$ is to the left of $V(P')$ and $v_k$ is in $V(P')$, or $v_j$ is in $V(P')$ and $v_k$ is to the right ot $V(P')$, or $v_j$ is to the left and $v_k$ is to the right of $V(P')$ and $D(v_j, v_k)$ does not touch $V(P')$ at all.

Suppose we have precomputed in $O(n^2)$ time, for every $v_i$ and $v_j$ the length $d_P(v_i, v_j)$ of the subpath of $P$ from $v_i$ to $v_j$. Consider any subpath $P'$ as above. We can create a graph $G_{P'}$ as follows. Let $P^l$ and $P^r$ be the subpaths of $P$ to the left and right of $P'$ respectively. Take $G$ and remove all incoming edges to nodes on $P^l$ except those in $P$ and all outgoing edges from nodes on $P^r$ except those in $P$. This will ensure that any path that we compute exiting $P^l$ doesn't reenter it and any path entering $P^r$ doesn't reexit it. Now remove all edges in $P'$ and split each node $v$ in $V(P')$ into two as before: a copy $v^{in}$ with all the remaining incoming edges and a copy $v^{out}$ with all the remaining outgoing edges. An example of the construction of $G_{P'}$ is given in Figure 2.

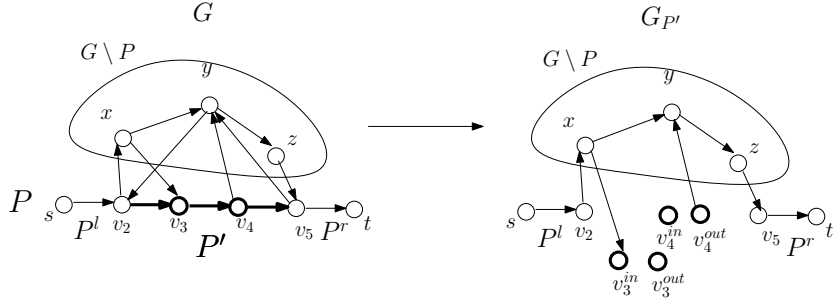

Figure 2: An example of the transformation of a graph $G$ into $G_{P'}$, given $P$ and subpath $P'$. Notice that edges $(y, v_2)$ and $(v_5, y)$ are removed, and that all nodes in $V(P')$ get split into two. The relevant paths in $G_{P'}$ corresponding to detour paths circumventing edges in $P$ are $s \to v_2 \to x \to v_3^{in}$, $s \to v_2 \to x \to y \to z \to v_5 \to t$ and $v_4^{out} \to y \to z \to v_5 \to t$.

Consider any (simple) path from $v_i^{out}$ to $v_j^{in}$ for $j > i$ and $v_i, v_j \in V(P')$. It corresponds to a detour in $G$ since it does not use any other nodes in $P$ by construction. Consider any path from $v_i^{out} \in V(P')$ to $t$ in $G_{P'}$. It corresponds to a portion of some replacement path for the edges of $P'$ after $v_i$. If we know the length of the shortest such path between $v_i^{out}$ and $t$, then we can compute the length of the shortest replacement path in $G$ that follows $P$ from $s$ to $v_i$ and circumvents all edges in $P'$ after $v_i$. Similarly, if we know the length of the shortest such path between $s$ and $v_i^{in}$ in $G_{P'}$, then we can compute the length of the shortest replacement path in $G$ that circumvents all edges in $P'$ before $v_i$ and follows $P$ from $v_i$ to $t$. Finally, the length of the shortest path between $s$ and $t$ in $G_{P'}$ is exactly the length of the shortest path in $G$ which circumvents all edges in $P'$. Thus, if we can compute the shortest paths in $G_{P'}$ between all pairs of nodes among the copies of nodes in $V(P')$, $s$ and $t$, then we can find the replacement paths for all edges in $P'$ in $O(|P'|^2 \log n)$ extra time with the successor search data structure approach from before. To compute these paths for $P'$ we use the following theorem by Yuster and Zwick [24].

**Theorem 5** (Yuster-Zwick [24]). *Given a directed graph $G$ with edge weights in $[-M, \ldots, M]$, one can compute in $\tilde{O}(Mn^\omega)$ time an $n \times n$ matrix $D$, so that the $(i, j)$ entry of the distance product $D \star D$ is the distance between nodes $i$ and $j$ in $G$.*

Since we only need to compute $O(|P'|^2)$ pairs of distances in $G_{P'}$, given the matrix $D$ for $G_{P'}$ from Theorem 5, we can do this in overall $\tilde{O}(Mn^\omega + n|P'|^2)$ time. If $|P'| = \sqrt{M} n^{(\omega-1)/2}$, then the runtime is $\tilde{O}(Mn^\omega)$.

This suggests the following bucketting approach: partition $P$ into $q \leq n^{(3-\omega)/2}/\sqrt{M}$ pieces $P_1, \ldots, P_q$ of size roughly $n/q = \sqrt{M} n^{(\omega-1)/2}$. Create a graph $G_{P_i}$ for each piece $i \in [q]$ as before. Compute the relevant shortest paths in overall time $O(qMn^\omega) = O(\sqrt{M} n^{(3+\omega)/2})$. After this, compute the lengths of all viable replacement paths in $O(n^2)$ extra time, and finally find the best replacement path for each edge using successor data structures in $O(n^2 \log n)$ time. The overall runtime of the algorithm is $O(\sqrt{M} n^{(3+\omega)/2}) = O(\sqrt{M} n^{2.688})$, slightly slower than the APSP-reduction technique in Algorithm 1.

## 3  A faster algorithm: Algorithm 3.

Here we will combine the above two algorithms to prove Theorem 2. The idea is as follows: we use some random sampling and Theorem 5 combined with fast matrix multiplication to compute partial distances between roughly equally spaced nodes on all detours. After this the bucketting technique of Algorithm 2 can actually be used to reduce the replacement paths problem to $n/b$ instances of APSP on graphs with $O(b)$ nodes and with only slightly increased edge weights. We can then use Zwick's APSP algorithm on the pieces and afterwards combine the results in roughly quadratic time. In the following we will ignore polylogarithmic factors in the runtime.

**Partitioning all long detours.** Call any detour on at least $n^{1-\varepsilon}$ edges *long*. Our algorithm will be able to handle short detours easily. Here we split the long detours into pieces which we will be able to handle just like the short detours. Following an idea from [21] we note that if we pick a set $B$ of size roughly $b = Cn^\varepsilon \ln n$ uniformly at random for large enough constant $C$, then every long detour path $D$ is partitioned by points of $B$ into pieces of size at most $n^{1-\varepsilon}$ with high probability.

**Lemma 1.** *Let $N \geq n$. Let $B$ be a random sample of $(C+3)n^\varepsilon \ln N$ nodes of $V$. With probability at least $1 - 1/N^C$, the nodes of $B$ touch all detours on at least $n^{1-\varepsilon}$ edges, so that the nodes of $B$ partition every detour into subpaths on at most $2n^{1-\varepsilon}$ edges each.*

*Proof.* Let $Q$ be a collection of $n^3$ paths, each on $n^{1-\varepsilon}$ nodes. We first show that the probability that $B$ contains at least one node from each path in $Q$ is at least $1 - 1/N^C$. We can assume that $B$ was formed by picking each node independently at random with probability $1/n$. Let $P \in Q$. The probability that none of the nodes of $B$ are in $P$ is $(1 - 1/n^\varepsilon)^{(C+3)n^\varepsilon \ln N} \leq 1/N^{C+3}$. The probability that there exists a path in $Q$ such that none of its nodes are in $B$ is $1/N^C$ by a union bound.

Now consider each detour $v_1 \to \ldots \to v_t$ on at least $n^{1-\varepsilon}$ nodes. Partition it by splitting off subpaths on $n^{1-\varepsilon}$ edges from one end until there is one path on $< n^{1-\varepsilon}$ nodes left. This creates $O(n^\varepsilon)$ disjoint subpaths on $n^{1-\varepsilon}$ nodes each and possibly one short path. Let $Q$ be the set of all long subpaths of all $O(n^2)$ long detours. There are at most $n^3$ of these and by our above argument, every one of them is hit by $B$ with high probability. Hence the number of edges between consecutive points of $B$ in each detour is at most $2n^{1-\varepsilon}$, and for any detour the number of edges from its beginning and the first point of $B$ on it and from its end and the last point of $B$ is also at most $2n^{1-\varepsilon}$. □

**Computing short detour subpaths.** We compute the matrix $D$ from Theorem 5 for the graph $G'$ used in Algorithm 1 (where the edges of $P$ are removed and the nodes are split into two). We only care about the distances between nodes of $B$ and nodes of $B \cup P$ in $G'$ which are of long detours, and about distances between nodes of $P$ which use short detours. All such distances are less than $O(Mn^{1-\varepsilon})$. Hence, we can disregard any entries of $D$ greater than $O(Mn^{1-\varepsilon})$. Then we can use the following theorem of Alon, Galil and Margalit [1] (following Yuval [25]).

**Theorem 6** ([1, 25]). *One can compute the distance product of two $n \times n$ matrices with entries in $\{-N, \ldots, N\}$ in $\tilde{O}(Nn^\omega) = O(Nn^{2.376})$ time.*

Using Theorem 6, we can compute the lengths $w(\cdot)$ of all short detour paths and all paths between nodes of $B$ and nodes of $B \cup P$ lying on long detours. The runtime is $O(Mn^{1-\varepsilon+\omega}) = O(Mn^{3.376-\varepsilon})$.

**Pasting the subpaths together to compute replacement paths.** We set aside the short detours. We process them in additional $O(n^2 \log n)$ time just as in Algorithm 1. Now we handle the rest of the detours. Since we know that the nodes of $B$ partition the long detours and we have computed all short paths between nodes of $B$ and nodes of $B \cup P$, we can focus on the graph which is only composed of $B \cup P$ with the new weights $w(\cdot)$ corresponding to short (on $\leq 2n^{1-\varepsilon}$ edges) pieces of long detours. All the weights are $O(Mn^{1-\varepsilon})$. Now we use the partitioning idea of Algorithm 2 to split the problem into $O(n/|B|)$ pieces each of size $O(|B|) = \tilde{O}(n^\varepsilon)$. If we use Algorithm 2 naively, even though $B$ is small, we would actually be left with a large part of $P$ in each subinstance, and the subinstance size would be linear in $n$. This can be fixed, however.

As in Algorithm 2, we partition $P$ into pieces $\{P_1, \ldots, P_{n/|B|}\}$ of size roughly $|B|$. We then create graphs $G_{P_i}$ for each piece. Consider a subinstance $G_{P_i}$. The nodes corresponding to nodes of $P$ are just the subpaths $P^l$ and $P^r$ to the left and right of $P_i$ with edges into $P^l$ and out of $P^r$ removed, together with in and out copies of the nodes of $P_i$. The number of nodes corresponding to $P_i$ is already $O(|B|)$. We will argue that we can replace $P^l$ and $P^r$ by $O(|B|)$ nodes each so that the edge weights of the graph increase by at most $O(Mn^{1-\varepsilon})$ and the distances between the relevant node pairs are preserved.

We will show how to replace $P^r$; $P^l$ can be processed similarly. Let $P^r = \{p_1, \ldots, p_t\}$ for $t \leq n$. Split $P^r$ into $t/c = O(n^\varepsilon)$ subpaths $\{(p_1 \to p_c), (p_{c+1} \to p_{2c}), \ldots, (p_{((t/c)-1)c+1} \to p_t)\}$ on $c - 1 = n^{1-\varepsilon}$ consecutive nodes. We will replace for each $i \geq 0$, the subpath $(p_{ic+1} \to p_{(i+1)c})$ by a new node $u_{i+1}$ so that for every node $b \in B$, the edge weight of $(b, u_{i+1})$ is $\min_{j=0}^{c-1} w(b, p_{ic+1+j}) + d_P(p_{ic+1+j}, p_{(i+1)c})$. As in Algorithm 1, there are no edges from any $u_i$ to $B$. Connect the $u_i$ nodes by a path by adding edges $(u_i, u_{i+1})$ with weight $d_P(p_{ic}, p_{(i+1)c})$ for all $1 \leq i < t/c$. See Figure 3 for an example transformation of a path $P^r$.

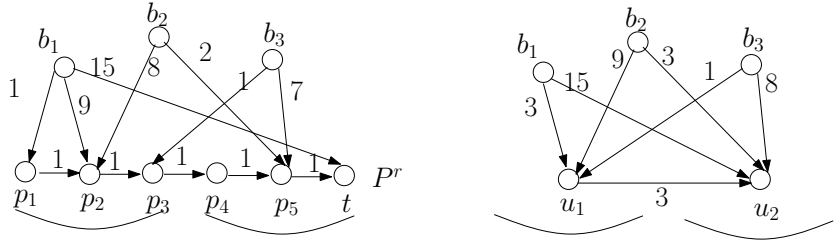

Figure 3: An example of the transformation of path $P^r$ in our main algorithm. Each consecutive subpath gets a node $u_i$ corresponding to the last node on the subpath. The nodes $u_i$ are chained in a path with weights corresponding to the distance of the last subpath nodes in the original path $P^r$. This is possible since any path that we are interested in never leaves $P^r$ once it enters it.

The weights of the edges in the new graph are still $O(Mn^{1-\varepsilon})$. The distance between any node in $P_i$ and $t$ is the same as before. The distance between $s$ and $t$ is the same as before. Assuming that all distances $d_P(x, y)$ are precomputed, the time to do this transformation is $O(n|B|)$ per piece $P_i$, and is thus $O(n^2)$ overall.

Thus, we have partitioned the problem in $\tilde{O}(n^2 + Mn^{3.376-\varepsilon})$ time into $O(n/|B|) = O(n^{1-\varepsilon})$ instances of APSP on graphs with $O(|B|) = \tilde{O}(n^\varepsilon)$ nodes with weights between $\{-N, \ldots, N\}$ where $N = O(Mn^{1-\varepsilon})$. In each subinstance, we can compute APSP using Zwick's algorithm [26] in time $O((Mn^{1-\varepsilon})^{0.681}(n^\varepsilon)^{2.575}) = O(M^{0.681}n^{0.681+1.894\varepsilon})$. The running time of this step over all $O(n^{1-\varepsilon})$ subinstances is asymptotically

$$n^{1-\varepsilon}[M^{0.681}n^{0.681+1.894\varepsilon}] = M^{0.681}n^{1.681+0.894\varepsilon}.$$

The overall runtime is
$$M^{0.681}n^{1.681+0.894\varepsilon} + Mn^{3.376-\varepsilon},$$
which is minimized when $M^{0.681}n^{1.681+0.894\varepsilon} = Mn^{3.376-\varepsilon}$, i.e. when $n^{1.894\varepsilon} = M^{0.319}n^{1.695}$. Then $n^\varepsilon = M^{0.168}n^{0.895}$, and the runtime is $O(M^{0.832}n^{2.481})$. We have proven Theorem 2.

## 4 An even faster algorithm.

In the previous section we reduced in $O(Mn^{1+\omega-\varepsilon})$ the replacement paths problem on $n$-node graphs with weights in $\{-M, \ldots, M\}$ to $n^{1-\varepsilon}$ instances of APSP in graphs on $n^\varepsilon$ nodes with weights in $\{-Mn^{1-\varepsilon}, \ldots, Mn^{1-\varepsilon}\}$. Here we make the observation that the APSP computation is not necessary, and we can instead solve the problem using recursion.

We will solve instances of the following problem.

**Definition 4.1** (Circumventing Paths Problem). *Let the following be given: $Z, n, p, M$, and $T = n/z^p$, a graph $\bar{G}$ on $O(T \log^p n)$ nodes with edge weights in $\{-N, \ldots, N\}$ with $N = O(MZ^p)$, an ordered list of $T$ tuples $((v_{i_1}^{in}, v_{i_1}^{out}), \ldots, (v_{i_T}^{in}, v_{i_T}^{out}))$, and for every $j \in [T]$, values $d_P(s, v_j)$ and $d_P(v_j, t)$. Every $v_{i_j}^{in}$ has only incoming edges and each $v_{i_j}^{out}$ has only outgoing edges. Compute for every $x \in [T]$:*

$$\min_{j,k:\ j \leq x < k} d_P(s, v_{i_j}) + d(v_{i_j}^{out}, v_{i_k}^{in}) + d_P(v_{i_k}, t),$$

*where $d(v_{i_j}^{out}, v_{i_k}^{in})$ is the distance in $\bar{G}$ between $v_{i_j}^{out}$ and $v_{i_k}^{in}$.*

Each ordered list of tuples above will correspond to a subpath $P'$ of $P$, and $d_P(u, v)$ is just the distance between $u$ and $v$ inside $P$. Now consider an instance of Circumventing Paths as in the definition and let $P'$ be a path $v_{i_1} \to \ldots \to v_{i_T}$ corresponding to the ordered list $((v_{i_1}^{in}, v_{i_1}^{out}), \ldots, (v_{i_T}^{in}, v_{i_T}^{out}))$. Consider a subpath $P_i$ of $P'$ on $T/Z = n/Z^{p+1}$ nodes. Let $B$ be a random sample of $\tilde{O}(T \log n)$ nodes from $\bar{G} \setminus P'$ so that all paths between pairs of nodes on $P_i$ on at least $Z$ nodes are partitioned by $B$ into pieces of length $O(Z)$, as in Lemma 1. Just as in Algorithm 3, use Theorems 5 and 6 and compute all distances shorter than $MZ^{p+1}$ between nodes of $\bar{G}$. This finds all detours on a small number of nodes. We can then compute the length of the paths in the original graph $G$ that these detours create and add the corresponding replacement paths to a global list $L$.

After all this computation, we can create two graphs $G(P_i)$ and $G^{P_i}$ for $P_i$ similar to Algorithm 3.

**The graph $G(P_i)$.** The graph $G(P_i)$ for $P_i$ is created as follows. First we take $P' \cup B$ and put an edge between any two nodes of $B$ with a weight corresponding to their distance in $\bar{G}$, if it was computed. We also add an edge between an out-node of $P_i$ and a node of $B$, or a node of $B$ and an in-node of $P_i$, using the distance (if it was computed) as a weight. The weights of the graph are now $O(MZ^{p+1})$.

As before, let $P^l$ and $P^r$ denote the subpaths of $P'$ to the left and right of $P_i$, respectively, where $P' = P^l \odot P_i \odot P^r$ ($\odot$ stands for concatenation). Abusing notation somewhat, let $P^l$ and $P^r$ also correspond to the copies of nodes of $P^l$ and $P^r$ which are in the ordered list of tuples of the Circumventing Paths instance.

We remove all in-copies of nodes in $P^l$ and all out-copies of nodes of $P^r$. Then, just as in Algorithm 3, we replace $P^l$ and $P^r$ each with a path on roughly $T/Z$ nodes, corresponding to subpaths on roughly $Z$ nodes each. We define new weights in these paths and between $B$ and the nodes of these paths according to distance, as in Figure 3. The main difference is that $v_{i_1}$ plays the role of $s$ and $v_{i_T}$ plays the role of $t$. More formally, we process $P^r$ as follows ($P^l$ is processed symmetrically): Let $P^r = \{p_1, \ldots, p_t\}$, where $p_t = v_{i_T}$. Split $P^r$ into $O(t/z) = O(T/Z)$ subpaths $\{(p_1 \to p_c), (p_{c+1} \to p_{2c}), \ldots, (p_{((t/c)-1)c+1} \to p_t)\}$ on $c - 1 = Z$ consecutive nodes. We will replace for each $i \geq 0$, the subpath $(p_{ic+1} \to p_{(i+1)c})$ by a new node $u_{i+1}$ so that for every node $b \in B$, the edge weight of $(b, u_{i+1})$ is $\min_{j=0}^{c-1} w(b, p_{ic+1+j}) + d_P(p_{ic+1+j}, p_{(i+1)c})$, where $w(\cdot, \cdot)$ are the distances computed in $\bar{G}$. Connect the $u_i$ nodes by a path by adding edges $(u_i, u_{i+1})$ with weight $d_P(p_{ic}, p_{(i+1)c})$ for all $1 \leq i < t/c$. This increases the maximum edge weight in the graph by at most $O(MZ^{p+1})$.

In each graph $G(P_i)$ we only need to compute:

1. the shortest path from the node corresponding to $v_{i_1}$ to all in-copies of nodes of $V(P_i)$;

2. the shortest path from all out-copies of nodes of $V(P_i)$ to the node corresponding to $v_{i_T}$;

3. the shortest path between the node corresponding to $v_{i_1}$ to the node corresponding to $v_{i_T}$;

4. for every $r \in [T/Z]$, the length $\ell$ of a path between $v_{i_j}^{out}$ and $v_{i_k}^{in}$ in $G(P_i)$ with $j \leq r < k$, such that the quantity $d_P(s, v_{i_j}) + \ell + d_P(v_{i_k}, t)$ is minimized over $j, k$ with $j \leq r < k$ such that $v_{i_j}, v_{i_k} \in P_i$.

1 and 2 compute the shortest pieces of detours one of the endpoints of which is not in $V(P_i)$. 3 computes an optimal detour circumventing all edges in $P_i$, and 4 is concerned with detours with endpoints within $V(P_i)$.

We note that one can compute 1, 2 and 3 in time $O((MZ^{p+1})(T \log n/Z)^\omega)$ using Yuster and Zwick's single source shortest paths algorithm [24]. After those distances are computed, one can convert them to lengths of candidate replacement paths as follows: Suppose $d$ is a distance computed between the node corresponding to $v_{i_1}$ and a node $v_{i_j} \in V(P_i)$. Then $d_P(s, v_{i_1}) + d + d_P(v_{i_j}, t)$ is the length of some replacement path which circumvents all edges of $P_i$ before $v_{i_j}$. Similarly, if $d$ is the distance computed between a node $v_{i_j} \in V(P_i)$ and the node corresponding to $v_{i_T}$, then $d_P(s, v_{i_j}) + d + d_P(v_{i_T}, t)$ is the length of some replacement path which circumvents all edges of $P_i$ before $v_{i_j}$.

**The graph $G^{P_i}$.** The graph $G^{P_i}$ is formed from $G(P_i)$ by removing the nodes corresponding to $P^l$ and $P^r$. Point 4 from above can be done with recursive calls to computing circumventing paths in the graph $G^{P_i}$.

Pseudocode for our approach is given in Algorithm 1 and Procedures 2 and 3. The intuition behind the algorithm is as follows. $G^{P'}$ gets much smaller than $G$ (some size $T = n/Z^i$ for a parameter $Z$), and there are at most $Z^i$ subinstances of that size. For each $i$, the subinstances correspond to a partition of $P$ into consecutive pieces of size roughly $n/Z^i$. Although each subinstance computes some detours which are *short* (only on $Z$ nodes), since the subinstance has weights which are large ($O(MZ^i)$), these short detours correspond to paths in $G$ of large weight.

After the computation of the short detours, the subinstance splits itself into a finer partition while increasing the maximum edge weight slightly in each new subinstance, and recurses on all the new subinstances. We show that the time due to the recursive calls is $\tilde{O}(ZMn^\omega)$, provided $Z = \Omega(\log^3 n)$. The number of circumventing paths which are computed as candidates for replacement paths is at most $O(n^2)$ and so the overall runtime is $\tilde{O}(Mn^\omega)$.

**Reminder of Theorem 3.** *The replacement paths problem for nodes $s$ and $t$ in an $n$-node graph with weights in $\{-M, \ldots, M\}$ can be solved by a randomized algorithm in $\tilde{O}(Mn^\omega)$ time if $\omega > 2$, or in $O(Mn^{2+\varepsilon})$ for every $\varepsilon > 0$ if $\omega = 2$.*

*Proof.* First compute the shortest path $P = \{s = v_1 \to \ldots, v_q = t\}$ between $s$ and $t$. Proceed as in Algorithm 1. Step 3 of Algorithm 1 takes $O(n^2)$ time. We will later show that $|L| = O(n^2)$. Because of this, the rest of the steps 6–16 of Algorithm 1 take $O(n^2 \log n)$ time.

Now consider Procedure 2. Consider level $i = 1, \ldots, \log n / \log Z$ of the recursion tree. It contains roughly $Z^i$ instances of size $O((n/Z^i) \log^i n)$. The weight size in any such instance is $O(MZ^i)$ For each of these, step 12 of Algorithm 2 takes $\tilde{O}((MZ^i)((n/Z^i) \log^i n)^\omega) = \tilde{O}((Mn^\omega/Z^{i\omega-1}) \log^{i\omega} n)$ time, and step 15 takes $\tilde{O}((MZ^{i+1})((n/Z^i) \log^i n)^\omega) = \tilde{O}((ZMn^\omega/Z^{i\omega-1}) \log^{i\omega} n)$ time.

Steps 16–19 add $O(n^2/Z^{2i})$ detour distances to $L$ for each of the $Z^i$ calls at the $i$th recusion tree level. The overall number of new items added to $L$ at level $i$ is $O(n^2/Z^i)$.

Consider Algorithm 3. For a call in the $j$th level of the recursion tree, and each subpath $P'_i$, steps 8–11 take $\tilde{O}((MZ^{j+1})((n/Z^{j+1}) \log^{j+1} n)^\omega)$ time via Yuster and Zwick's algorithm for SSSP [24]. The number of calls to steps 8–11 at the $j$th level of the recursion tree is $Z^{j+1}$ so the runtime is

$$M(n^\omega \log^{(j+1)\omega} n)/Z^{(j+1)(\omega-2)}.$$

The number of new detours added to $L$ in steps 8–15 is $O(n)$ for every level of the recursion tree. The number of levels of the recursion tree is $O(\log n / \log Z)$. The overall number of items in $L$ at the end of all recursive calls is hence asymptotically

$$n \log n / \log Z + \sum_{i=0}^{\log n / \log Z} (n^2/Z^i).$$

This is $O(n^2)$ for any $Z \geq 2$, say.

The overall running time is asymptotically

$$n^2 \log n + \sum_{i=0}^{\log n / \log Z} \frac{ZMn^\omega \log^{i\omega} n}{Z^{i(\omega-1)}} + \frac{Mn^\omega \log^{(i+1)\omega} n}{Z^{(i+1)(\omega-2)}}.$$

If $\omega > 2$, the above runtime is $\tilde{O}(ZMn^\omega)$, say for any $Z \geq 2 \log^{\omega/(\omega-2)} n$. Hence picking $Z = \Theta(\log^{\omega/(\omega-2)} n)$ we obtain a runtime of $\tilde{O}(Mn^\omega)$. If $\omega = 2$, then we can pick $Z = n^\varepsilon$ for any $\varepsilon > 0$. The runtime becomes $O(Mn^{2+2\varepsilon})$.

We need to make sure that all random samples $B$ work with high probability. The number of random samples $B$ which are picked is at most $O(n \log n)$ and the probability that a fixed one of them doesn't work is $1/\operatorname{poly}(n)$, so by a union bound, the probability that all of them work is at least $1 - 1/\operatorname{poly}(n)$. □

**Algorithm 1** ReplacementPaths

1: Input: $G = (V, E)$, $s$, $t$; Global: $L$
2: Compute the shortest path $P = s = v_1 \to v_2 \to \ldots, v_l = t$ using Yuster-Zwick's algorithm [24]
3: Compute $\forall v_i, v_j \in P$ their distance $d_P(v_i, v_j)$ in $P$ via dynamic programming
4: Create a graph $G'$ by removing edges on $P$ and splitting every node $v \in P$ into $\{v^{in}, v^{out}\}$
5: Call RECURSIVE($G'$, $P$, $s$, $t$, $M$)
6: Sort all elements $(v_j, v_k, d) \in L$ in nondecreasing order of $d$
7: Store every $(v_i, i)$ in a binary search tree $T$
8: **while** $L \neq \emptyset$ **do**
9:    Pop $(v_j, v_k, d)$ from $L$
10:   Find successor $v_q$ of $v_j$ in $T$
11:   **for** every $(v_p, p) \in T$ with $i \leq p \leq j$ **do**
12:      Set $d[v_p, v_{p+1}] = d$
13:      Remove $v_p$ from $T$
14:   **end for**
15: **end while**
16: Output $d[\cdot, \cdot]$

**Procedure 2** RECURSIVE

1: Input: $(G', P', u, v, N)$, Global: $L, P, d_P(\cdot, \cdot)$
2:
3: **if** $G'$ has size $O(Z \log n)$ **then**
4:   Compute the distances $d_{G'}(\cdot, \cdot)$ between all pairs of nodes in $P'$ in $G'$
5:   **for** every $v_i, v_j \in P', i < j$ **do**
6:     Compute $d = d_P(s, v_i) + d_{G'}(v_i, v_j) + d_P(v_j, t)$
7:     Add $(v_i, v_j, d)$ to $L$
8:     Return
9:   **end for**
10: **end if**
11:
12: Compute matrix $D$ so that $(D \star D)[u, v]$ is the distance of $u$ and $v$ in $G'$
13: Sample a set $B$ of $O((|G'| \log n)/Z)$ nodes from $G' \setminus P'$
14: Replace all entries $D[u, v] \geq N \cdot Z$ with $D[u, v] = \infty$.
15: Compute $D \star D$
16: **for** all $v_i, v_j \in P'$ with $i < j$ and $(D \star D)[v_i, v_j] < \infty$ **do**
17:   Compute $d = d_P(s, v_i) + (D \star D)[v_i, v_j] + d_P(v_j, t)$
18:   Add $(v_i, v_j, d)$ to $L$
19: **end for**
20:
21: Call ProcessSubpath($G', P', u, v, B, N$)

**Procedure 3** ProcessSubpath

1: Input: $(G', P', u, v, B, N)$, Global: $L, P, d_P(\cdot, \cdot)$
2:
3: Split $P'$ into $Z$ consecutive buckets $\{P'_1, \ldots, P'_Z\}$ of size $|P'|/Z$
4: Remove all nodes in $G' \setminus \{B \cup P'\}$ from $G'$
5: **for** each $P'_i = u'_i \to \ldots \to v'_i$ **do**
6:    Let $P^r$ and $P^l$ consist of the nodes to the right and left of $P'_i$, respectively
7:    Replace $P^r$ and $P^l$ by paths on $|P'|/Z$ nodes as in Figure 3 creating graph $G'_i$
8:    Find the distance $d_i(u, v)$ between $u$ and $v$ in $G'_i$
9:    Add $(u'_i, v'_i, d_P(s, u) + d_i(u, v) + d_P(v, t))$ to $L$
10:    Find the distances $d_i(u, v_j)$ from $u$ to all nodes $v_j \in P'_i$ in $G'_i$
11:    Find the distances $d_i(v_j, v)$ from all nodes $v_j \in P'_i$ in $G'_i$ to $v$
12:    **for** all $v_j \in P'_i$ **do**
13:      Compute $d_u = d_P(s, u) + d_i(u, v_j) + d_P(v_j, t)$
14:      Add $(u'_i, v_j, d_u)$ to $L$
15:      Compute $d_v = d_P(s, v_j) + d_i(v_j, v) + d_P(v, t)$
16:      Add $(v_j, v'_i, d_u)$ to $L$
17:    **end for**
18:    Remove all nodes corresponding to $P' \setminus P'_i$ from $G'_i$
19:    Call RECURSIVE$(G'_i, P'_i, u'_i, v'_i, NZ)$
20: **end for**